\documentclass[aps,pra,reprint,longbibliography,superscriptaddress,nofootinbib]{revtex4-1}

\usepackage{amsmath,amssymb,MnSymbol,graphicx,amsfonts,csquotes,hyperref}

\hypersetup{
colorlinks=true,       
    linkcolor=blue,          
    citecolor=blue,        
    urlcolor=blue }

\begin{document}

\title{Practical Advantages of Almost-Balanced-Weak-Values Metrological Techniques}

\author{Juli\'{a}n Mart\'{i}nez-Rinc\'{o}n}
\email{jmarti41@ur.rochester.edu}
\affiliation{Department of Physics and Astronomy, University of Rochester, Rochester, New York 14627, USA}
\affiliation{Center for Coherence and Quantum Optics, University of Rochester, Rochester, New York 14627, USA}
\author{Zekai Chen}
\affiliation{Department of Physics and Astronomy, University of Rochester, Rochester, New York 14627, USA}
\affiliation{Center for Coherence and Quantum Optics, University of Rochester, Rochester, New York 14627, USA}
\author{John C. Howell}
\affiliation{Department of Physics and Astronomy, University of Rochester, Rochester, New York 14627, USA}
\affiliation{Center for Coherence and Quantum Optics, University of Rochester, Rochester, New York 14627, USA}
\affiliation{Institute of Optics, University of Rochester, Rochester, New York 14627, USA}
\affiliation{Institute for Quantum Studies, Chapman University, Orange, California 92866, USA}

\date{\today}

\begin{abstract}
Precision measurements of ultra-small linear velocities of one of the mirrors in a Michelson interferometer are performed using two different weak-values techniques. We show that the technique of Almost-Balanced Weak Values (ABWV) offers practical advantages over the technique of Weak-Value Amplification (WVA), resulting in larger signal-to-noise ratios and the possibility of longer integration times due to robustness to slow drifts. As an example of the performance of the ABWV protocol we report a velocity sensitivity of 60 fm/s after 40 hours of integration time. The sensitivity of the Doppler shift due to the moving mirror is of 150 nHz. 
\end{abstract}

\maketitle

\section{Introduction}
Post-selected weak measurements have proven to be useful for metrology of parameter estimation during the recent years~\cite{ReviewNori,ReviewSvensson}. These techniques consist of weakly coupling a system to a meter and then measuring the meter only when a successful post-selection on the system occurs. Weak-Value Amplification (WVA) is one such technique, where strong discarding of data is part of the requirements to induce a large signal proportional to the parameter of interest. The technique was initially proposed almost thirty years ago~\citep{AAV}, and has been extensively studied and applied after the first successful implementation twenty years later~\cite{Hosten}. The technique has been used to measure shifts of a laser frequency~\cite{David}, linear velocities~\cite{Velocimetry}, optical phases~\cite{LuisJose,LuisJoseDelays}, displacements due to the optical spin Hall effect~\cite{Hosten,ModifiedHallEffect,HallEffectNanometal,Jayaswal,Pfeifer}, temperature shifts~\cite{LuisJoseTemperature,Thermostat}, angular rotations of a laser beam~\cite{AngularMomentum}, tilts of a mirror~\cite{Dixon,Starling,Kasevich,Turner,Concatenated}, polarization rotations~\cite{First}, angular rotations of chiral molecules~\cite{ChiralJoe,Chiral2016}, and glucose concentration~\cite{Glucose}. Advantages of WVA for metrology rely on practical aspects of anomalous amplification and low detected power. These aspects allow one to increase the signal-to-noise ratio in technical noise-limited scenarios. For a review on WVA and its advantages see Refs.~\cite{ReviewShikano,Understanding,TorresWVA,Brunner,QAverage,Defeats,Pang2,Gabriel,PRX,NoiseExp,WVASystematicErrors,Harris2016}.

Extensions or similar approaches to WVA have also been proposed. For example, it has been shown that the postselection probability distribution adds useful information to the parameter estimation task~\cite{Gabriel2016}. There is also a second weak-values technique, known as Inverse Weak Value, where the post-selection in the system induces a stronger back-action in the meter than the weak system-meter coupling~\cite{inverse,ReviewNori}. Precision measurements of phase~\cite{inverse} and tilts~\cite{TiltsInverse} in Sagnac interferometric configurations have been reported using such a protocol. Taking a different approach, Str\"{u}bi and Bruder proposed the use of two detectors to collect all of the information under a different post-selection procedure than WVA~\cite{Strubi}.    Experimental demonstration of the robustness \textquote{against not only misalignment errors but also the wavelength dependence of the optical components} of such a protocol was soon demonstrated~\cite{ExpJointWM}. It was also shown than a WVA-like response can be obtained in the difference signal of the two-detector protocol~\cite{ABWV}. Simulating an anomalous amplification in a Homodyne detection procedure, the technique has been dubbed Almost-Balanced Weak Values (ABWV). The ABWV technique has been used to measure angular velocities of the linear polarization of laser pulses with a precision of 22 nrad/s after $\sim$11 hours of integration time~\cite{ABWV}, and more recently, to measure angular velocities of a rotating mirror with a precision of $\sim$4.9 nrad/s with one minute of collection time~\cite{Weitao}. 

WVA has been successfully used to measure ultra-small linear velocities 
of a moving mirror in a Michelson interferometer on a table-top  configuration~\cite{Velocimetry}. The best reported result is of a velocity of 400 fm/s (or 1 $\mu$Hz Doppler shift) after averaging for a little longer than two hours. Lack of robustness to long drifts did not allow for longer integration times. We evaluate here the performance of the ABWV technique to carry out the same metrological task, and find it superior to the WVA case. We report a sensitivity of 60 fm/s (or 150 nHz Doppler shift) after 40 hours of collection time.

The ABWV protocol has proven to offer larger amplification than the WVA approach~\cite{Weitao}, however it is still an open question if it offers noise-mitigation advantages or not. We perform the velocities measurements under equal conditions for both techniques (WVA and ABWV) and for six different frequencies on the driving mirror. We show that ABWV offers on average twice better signal-to-noise ratio than WVA for measurements of linear velocities.

In Section~\ref{section theory} we introduce the experiment used to measure the linear velocities, and how the comparison between both techniques is done. In Section~\ref{section exp} we reveal details and results of the experiments, to finally discuss them in Section~\ref{section conclusions}. 
 
\section{Weak-value vs Almost-Balanced-Weak-Values Amplification}\label{section theory}

\begin{figure}[htbp]
\centering
\includegraphics[width=\linewidth]{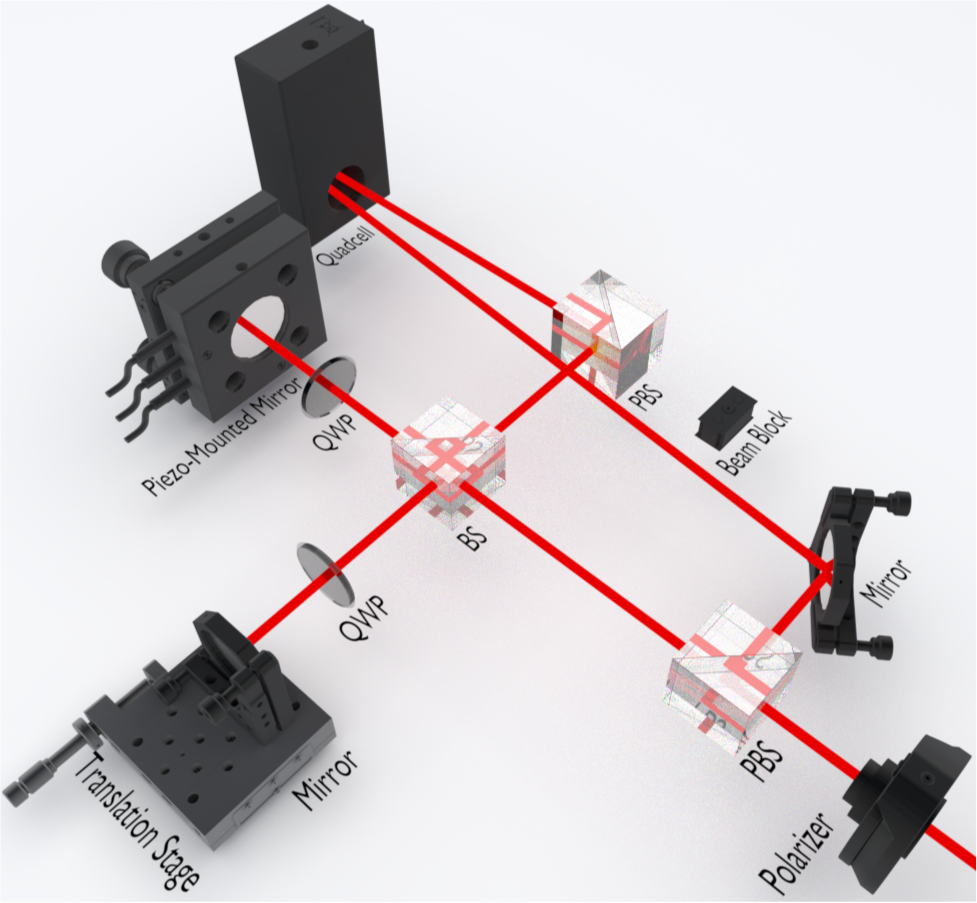}
\caption{Schematic of the experimental setup (not to scale) to measure linear velocities of a piezo-mounted mirror. Horizontally polarized laser pulses are sent to a Michelson interferometer. Quarter-wave-plates (QWP) and polarized beam-splitters (PBS) are used to send the vertically-polarized output pulses to a balanced detector. A beam-block is used to allow collection of both or just one of the ports.}
\label{fig:setup}
\end{figure}

We are interested in estimating the linear velocity of one of the mirrors in a Michelson interferometer, as sketched in Fig.~\ref{fig:setup}. The interferometer consists of one non-polarizing beam splitter and two mirrors, with one of them on a piezo-driven mount moving at a constant ultra-small speed. Phase noise in the interferometer is minimized by mounting the second mirror on a translation stage. This design allows for control of the arms' length difference. The input polarization of the laser pulses is set to horizontal. Using a quarter-wave-plate in each arm of the interferometer both vertically-polarized outputs can be collected using polarized beam splitters. The two optical outputs are sent to a balanced detector, where the difference and the sum electrical signals are recorded.

Pulses, with a power distribution given as $P_{in}(t)=P_0e^{-t^2/2\tau^2}$, are sent to the interferometer. $P_0$ is the peak power and $\tau$ is the characteristic length of the pulse. The piezo-mounted mirror can be precisely controlled to set the optical outputs in a bright/dark or an almost balanced configuration. The WVA configuration is obtained by tracking only the dark port. This arrangement is done by setting one output beam as the bright port and then blocking it to avoid its detection. The piezo is used to control the dark port, which takes the following power distribution,
\begin{eqnarray}\label{WVA}
P^{WVA}(t)&=&\frac{1}{4}\left|\beta_1-e^{i(\epsilon+2kvt)}\right|^2P_{in}(t)\nonumber\\
&=&P_0\left[\left(\frac{1-\beta_1}{2}\right)^2+\beta_1\sin^2(\epsilon/2+kvt)\right]e^{-t^2/2\tau^2},
\end{eqnarray}
where $\epsilon$ is a tunable phase difference between paths, $v$ is the velocity of the piezo-driven mirror, and $k=2\pi/\lambda$ is the wave number. The parameter $\beta_1$ is the relative transmission amplitude between both arms of the interferometer at the given port. This parameter defines the visibility of the dark port as $2\beta_1/(1+\beta_1^2)$. A value of $\beta_1$ smaller than unity accounts for imperfections of the optical elements in the experiment~\cite{Concatenated}. 
 
The usual WVA approximation is obtained for weak interactions and small postselection angles, i.e. $kv\tau\ll\epsilon/2\ll1$. For such a limit, in a perfect interferometer ($\beta_1=1$), the power distribution at the dark port takes the form $\sim P_0\sin^2(\epsilon/2)e^{-(t-\delta t)^2/2\tau^2}$ where $\delta t=4kv\tau^2/\epsilon$~\cite{Velocimetry}. The stronger the discarding of data counts (smaller $\epsilon$) the larger the induced time shift $\delta t$ in the pulse. We consider here only the weak interaction approximation, i.e. $kv\tau\ll1$, and evaluate the performance of the interferometer as a function of the phase $\epsilon$. Eq.~(\ref{WVA}) is then approximated as a Gaussian distribution with amplitude
\begin{equation}
P^{WVA}_{peak}\approx P_0\left[\left(\frac{1-\beta_1}{2}\right)^2+\beta_1\sin^2(\epsilon/2)\right],
\end{equation}
and time shift
\begin{equation}\label{shiftWVA}
\delta t^{WVA}\approx\frac{4\beta_1kv\tau^2\sin(\epsilon)}{(1-\beta_1)^2+4\beta_1\sin^2(\epsilon/2)}.
\end{equation}
  
The ABWV technique requires the use of both output ports in the interferometer. For this case, the beam-block is removed and no discarding of data counts occurs (as it is explicitly shown in Fig.~\ref{fig:setup}). The almost-balanced configuration is set by moving the piezo-driven mirror an extra distance $\lambda/8$. The power distributions of the two ports take the form
\begin{eqnarray}
P^{ABWV}_{1,2}(t)&=&\frac{1}{4}\left|\beta_{1,2}\mp e^{i(\epsilon+2kvt+2k(\lambda/8))}\right|^2P_{in}(t)\nonumber\\
&=&\left[\frac{1+\beta_{1,2}^2}{4}\pm\frac{\beta_{1,2}}{2}\sin(\epsilon+2kvt)\right]P_{in}(t),\nonumber
\end{eqnarray}
where we have assumed that both ports have different visibilities, i.e. $\beta_2\neq\beta_1$. 

The sum signal takes the form
\begin{eqnarray}
P^{ABWV}_{sum}&=&P^{ABWV}_1+P^{ABWV}_2\label{sum}\\&=&P_0\left[\frac{2+\beta_1^2+\beta_2^2}{4}+\frac{\beta_1-\beta_2}{2}\sin(\epsilon+2kvt)\right]e^{-t^2/2\tau^2},\nonumber
\end{eqnarray}
and the difference signal
\begin{eqnarray}
P^{ABWV}_{diff}&=&P^{ABWV}_1-P^{ABWV}_2\label{diff}\\&=&P_0\left(\frac{\beta_1+\beta_2}{2}\right)\left[\frac{\beta_1-\beta_2}{2}+\sin(\epsilon+2kvt)\right]e^{-t^2/2\tau^2}.\nonumber
\end{eqnarray}

These expressions reduce to the ones in Ref.~\cite{ABWV} by assuming a perfect interferometer, $\beta_1=\beta_2=1$. In addition, if the weak-value approximation is considered, $kv\tau\ll\epsilon/2\ll1$, the expressions take the simple form $P_{sum}^{ABWV}\approx P_{in}(t)=P_0e^{-t^2/2\tau^2}$ and $P_{diff}^{ABWV}\approx P_0\sin(\epsilon)e^{-[t-2kv\tau^2/\epsilon]^2/2\tau^2}$. In our experiment $\beta_1\approx\beta_2\approx0.7$. We use equations (\ref{sum}) and (\ref{diff}) to evaluate the performance of the technique as a function of $\epsilon$ under the weak interaction approach, $kv\tau\ll1$. The effective time shift between the sum and the difference signal is given by
\begin{equation}\label{shiftABWV}
\delta t^{ABWV}\approx\frac{8(1+\beta_1\beta_2)kv\tau^2\cos\epsilon}{[2\sin\epsilon+\beta_1-\beta_2][2+\beta_1^2+\beta_2^2+2(\beta_1-\beta_2)\sin\epsilon]}.
\end{equation}

We are interested on comparing the performance of both techniques to induce time shifts (Eqs.~\ref{shiftWVA} and~\ref{shiftABWV}), and the  corresponding estimates of the velocity $v$. 

\section{Experimental Results}\label{section exp}
A 795-nm continuous-wave laser beam (Vescent Photonics distributed Bragg reflector laser diode D2-100-DBR) was sent through an Acousto-Optic Modulator (AOM). The laser and the AOM are not shown in Fig.~\ref{fig:setup}. The AOM was used to modulate a Gaussian profile in the field's distribution, i.e. $P_{in}(t)=P_0\exp\left(-t^2/2\tau^2\right)$, in the first-order diffracted beam, which was later coupled to a single mode patch cable. The non-Fourier band-limited pulses with repetition rate $f_r$ were launched and prepared with horizontal polarization before passing through a polarizing beam-splitter and sent to the interferometer. 

A piezo-driven mirror mount (Thorlabs KC1-PZ) was used for one of the mirrors to control the phase $\epsilon$ in the interferometer and to induce the constant linear speed $v$. Each quarter-wave-plate in the arms of the interferometer was set to $45^\circ$ with respect to the horizontal input light to make the outputs vertically polarized. The second mirror in the interferometer was mounted on a linear translation stage to make the arms' length difference no larger than tens of microns. This optimization, to reduce phase noise, was done by sweeping the laser frequency within a range of about 4 GHz and minimizing the phase readout of the interferometer by moving the stage. Each arm in the interferometer was about 4 cm long. 

The two output pulses of the interferometer were directed to two of the four detectors of a quadrant cell photoreceiver (Newport 2921). The output electrical signals were proportional to the sum and difference laser-power distributions. By blocking or unblocking one of the two optical ports and controlling the phase $\epsilon$ the system resembled either the WVA or the ABWV technique respectively~\footnote{In the case of ABWV, both sum and difference signals were used for data processing. In the case of WVA, only one optical port was measured, so $P_{sum}^{WVA}=P_{diff}^{WVA}=P_1^{WVA}=P^{WVA}$.}. The difference and sum electrical signals were directly recorded using an oscilloscope and a computer. \textit{No frequency filters nor lock-in amplifiers were used}, meaning that both techniques (WVA and ABWV) were compared under the same technical-limited conditions. 

A 60$\%$ duty-cycle triangle ramp with peak-to-peak voltage $V_{pp}=75$ mV and frequency $f_r$ was applied to the piezo actuators on the mirror mount. The linear velocity of the mirror during the positive ramp was given by $v=5\alpha V_{pp}f_r/3\sim (6.66$ nm)$\times f_r$, where $\alpha\approx53.33$ nm/V is the manual-given mount response. Both techniques were compared for six different frequencies ($f_r=0.5,1,2,5,10,20$ Hz) and for different values of $\epsilon$. For a given frequency $f_r$ and a given phase $\epsilon$, 118 pulses were used and recorded during the experiment. Each collected pulse consisted of 1250 data points and was numerically fitted to one of the distributions~(\ref{WVA}),~(\ref{sum}), or~(\ref{diff}). Such a fitting averages out fluctuations much faster than $f_r$ in the time-dependent pulse intensity. In addition, the fitting-obtained time shifts of 118 pulses were averaged for each couple $f_r$ and $\epsilon$.  

\begin{figure*}[htbp]
\centering
\includegraphics[width=1.00\linewidth]{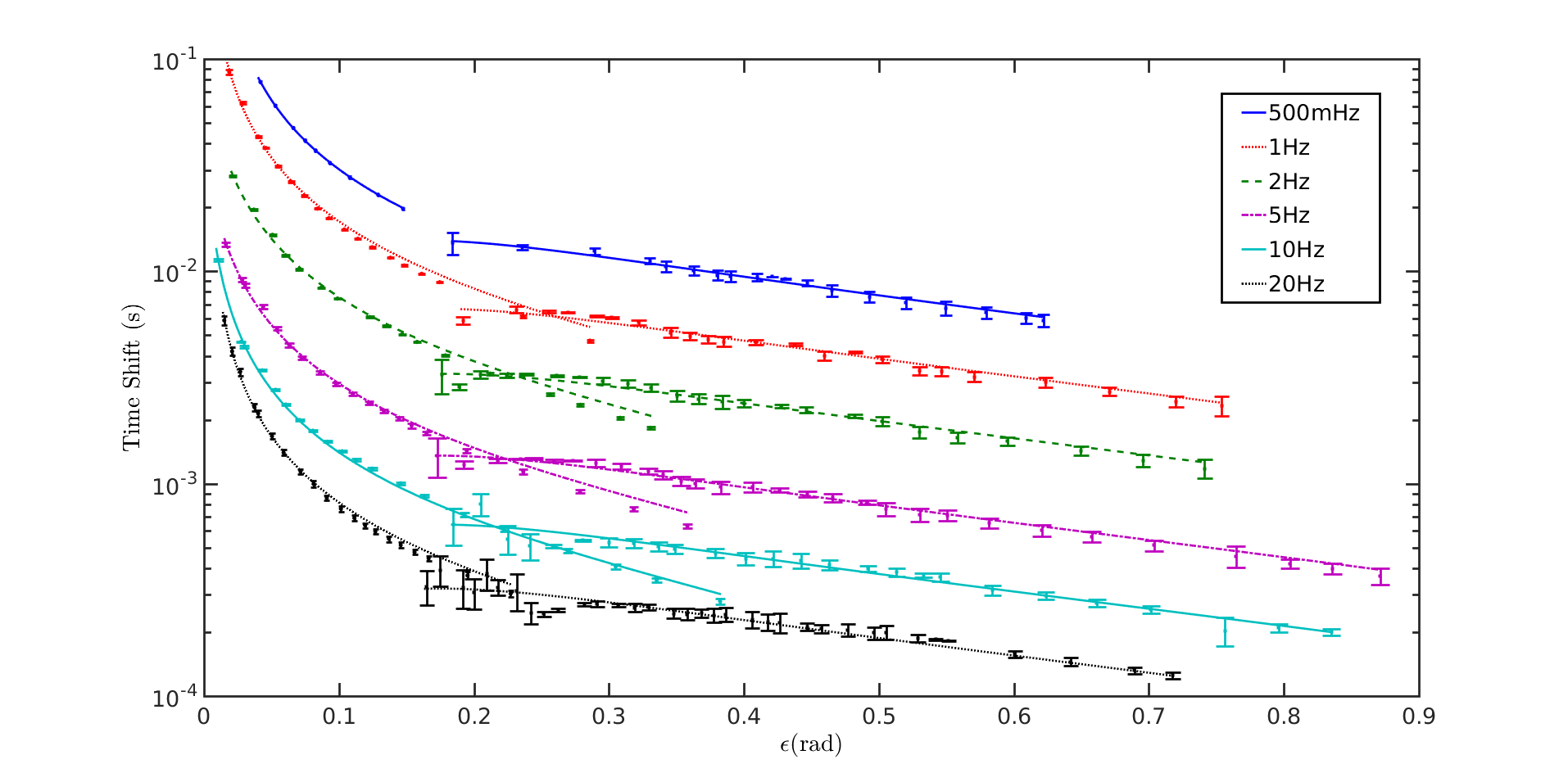}
\caption{Experimental estimates of the time shifts $\delta t^{WVA}$ (right side curves) and $\delta t^{ABWV}$ (left side curves). Both techniques are compared for six different frequencies $f_r$ and for experimentally-allowed values of $\epsilon$ in each case. From top to bottom the curves correspond to frequencies $f_r=$ 0.5, 1, 2, 5, 10, and 20 Hz.}
\label{fig:shifts}
\end{figure*}

Fig.~\ref{fig:shifts} shows the result of the estimated average time shifts for both techniques. In the case of WVA, each collected pulse was numerically fitted to distribution~(\ref{WVA}) setting $\beta_1$, $\tau$, $\epsilon$, and $v$ as free parameters. Correction to laser power drift was done to each pulse before doing the numerical fitting. The obtained values were then used to estimate the time shift $\delta t^{WVA}$ in Eq.~(\ref{shiftWVA}). These results are shown as the six curves on the right side of Fig.~\ref{fig:shifts}. For a given $f_r$ each data point corresponds to the average of 118 time shifts for each value of $\epsilon$, and the curve represents a (second) numerical fitting of these data points to Eq.~(\ref{shiftWVA}). Such a curve is introduced for eye-guiding purposes. For the case of ABWV, two distributions (sum and difference) were recorded for each of the 118 pulses for a given frequency $f_r$ and a given phase $\epsilon$. Each couple was numerically fitted to equations~(\ref{sum}) and~(\ref{diff}), adding the extra free parameter $\beta_2$. Power drift correction was not necessary in this case. The values obtained from the fitting for the five parameters were used to evaluate the time shift $\delta t^{ABWV}$ using Eq.~(\ref{shiftABWV}). These shifts are shown in the six curves on the left side of Fig.~\ref{fig:shifts}.

In Fig.~\ref{fig:shifts}, we notice the following: first, the ABWV technique offers larger amplification (i.e. larger time shifts) than WVA, and second, the amplification monotonically grows for small values of $\epsilon$ (WVA breaks down for $\epsilon\lesssim 0.2$ rad). The former was noted when the technique was originally proposed~\cite{ABWV}, and the latter was subsequently experimentally demonstrated~\cite{Weitao}. For a given frequency $f_r$ (and velocity $v$), the smaller the phase $\epsilon$ the larger the time shift. The maximum amplification in the ABWV case was on average 13 times larger than the optimal case in WVA. The lowest amplification ratio was 5.7 for $f_r=500$ mHz and the largest was 21 for $f_r=10$ Hz. These amplification gain differences are due to the experimentally-allowed minimum values for $\epsilon$ obtained in the ABWV case for each frequency $f_r$.

\begin{figure}
\centering
\includegraphics[width=\linewidth]{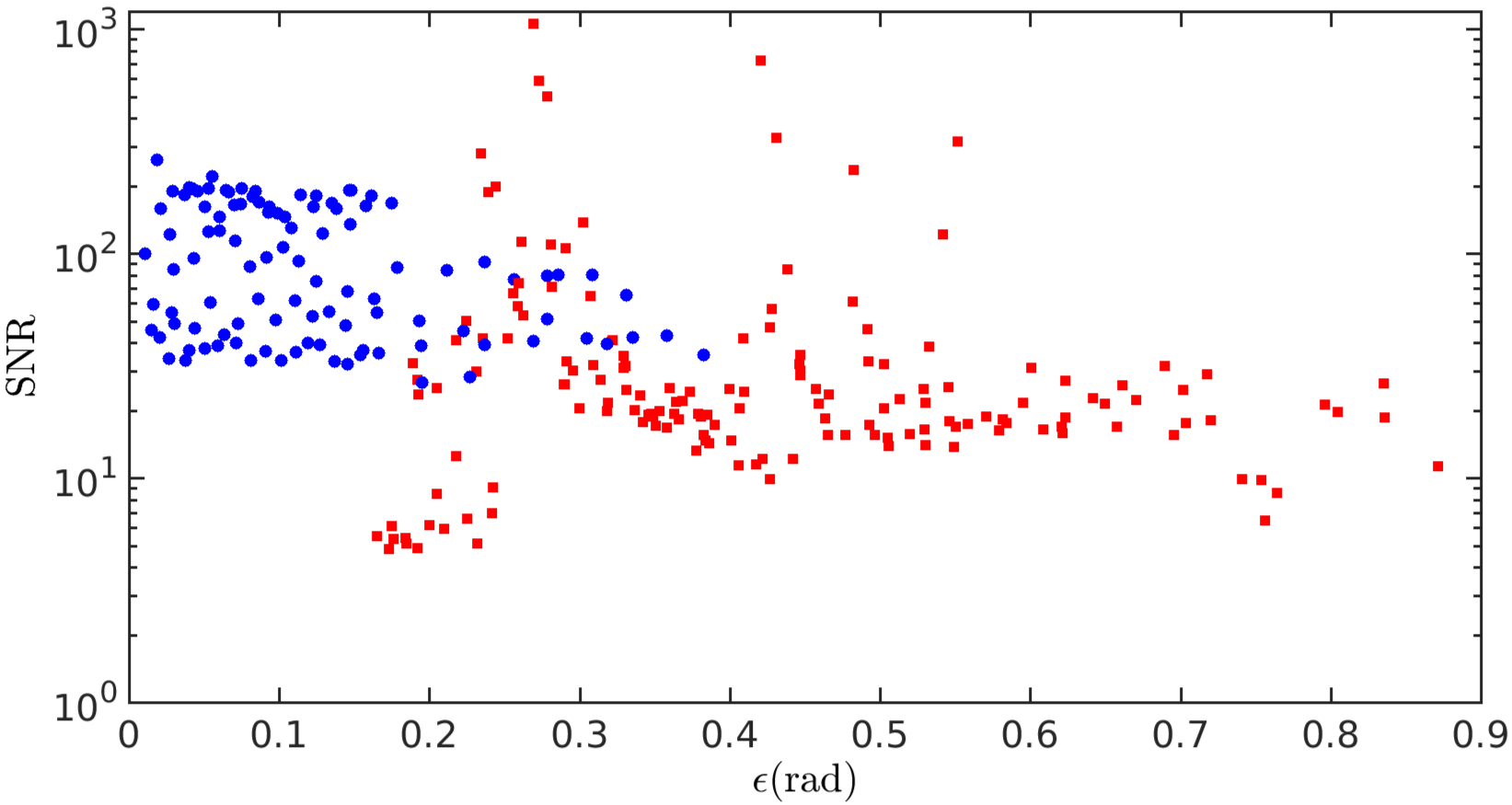}
\caption{Signal-to-Noise Ratio (SNR) results obtained from Fig.~\ref{fig:shifts}. Values for the ABWV (WVA) case are represented as blue circles (red squares).}
\label{fig:SNR}
\end{figure}

It is also shown in Fig.~\ref{fig:shifts} that the ABWV technique shows a relatively consistent precision (size of error bars) for all data points. This behaviour is not seen for the WVA technique, as it is explicitly shown in Fig.~\ref{fig:SNR}. The ABWV case (blue circles) offers in average $\sim1.8$ larger signal-to-noise ratio (SNR) than the WVA case (red squares). The larger the frequency $f_r$ the larger the SNR, however we compare in Fig.~\ref{fig:SNR} the overall behaviour of both techniques. As an exception of few data points, the ABWV case offers much better SNR performance than the WVA case.

The consistency of the SNR results of the ABWV technique over the WVA technique relies on technical details that makes the implementation of the almost-balanced case advantageous. These advantages are:
\begin{itemize}

\item The ABWV protocol is more robust to slow fluctuations of the input laser power than the WVA one. This is due to the fact that $P_0$ must be independently measured in the WVA case before setting the interferometer in the dark port. For the ABWV case, we approximated Eq.~(\ref{sum}) as $P_{sum}^{ABWV}\approx P_0e^{-t^2/2\tau^2}$ since $\beta_1\approx\beta_2\approx0.7$, and obtained real-time values of $P_0$ and $\tau$ for each pulse. This fact alone allows for longer collection times in the ABWV case. In order to improve the estimates for the WVA case, we separately estimated the power drift of the 118 pulses collected for each data point in Fig.~\ref{fig:shifts}. We then corrected for the drift in Eq.~(\ref{WVA}) before running the numerical fittings. The drift-subtracted results for WVA are not better than the raw fittings for ABWV, as shown in Fig.~\ref{fig:SNR}.

\item Due to the robustness to slow drifts in the input power of the ABWV technique, the estimation task is robust to slow drifts in the interferometer's alignment. After obtaining values of $P_0$ and $\tau$ for each pulse from fittings to $P_{sum}^{ABWV}\approx P_0e^{-t^2/2\tau^2}$, a systematic error-free estimation of $\beta_1$, $\beta_2$, $\epsilon$, and $v$ were performed from numerical fittings to the difference signal in  Eq.~(\ref{diff}). 

\item The ABWV design offers better repeatability of the experimental results. This behaviour can be observed in the values obtained for the parameters $\tau$, $\beta_1$, and $\beta_2$ from the numerical fittings. Table~\ref{table:one} shows that the ABWV technique allows for lower deviation on the estimations of these parameters. Since the ABWV protocol removes background and common noise by differencing, estimates of $\beta_1$ and $\beta_2$ are more accurate than independently using the WVA protocol for each port.

\end{itemize}

\begin{table}
\begin{tabular}{| c | c | c | c | c |}
\hline
& $f_r$ (Hz) & $\tau$ (ms) & $\beta_1$ & $\beta_2$\\ \hline
WVA & 0.5 & $299\pm5$ & $0.72\pm0.01$ & - \\
 & 1 & $149.0\pm0.2$ & $0.72\pm0.02$ & - \\
 & 2 & $74.6\pm0.1$ & $0.72\pm0.02$ & -\\
  & 5 & $29.84\pm0.05$ & $0.71\pm0.02$& - \\
 & 10 & $14.90\pm0.02$ &  $0.71\pm0.02$ & - \\
 & 20 & $7.45\pm0.02$ &  $0.72\pm0.02$ & - \\ \hline
 ABWV & 0.5 & $298.2\pm0.2$ & $0.695\pm 0.002$ & $0.686\pm 0.002$ \\
 & 1 & $149.08\pm0.05$ & $0.693\pm 0.003$ & $0.689\pm 0.003$ \\
 & 2 & $74.56\pm0.02$ & $0.682\pm 0.003$ & $0.700\pm 0.003$ \\
  & 5 & $29.831\pm0.006$ & $0.683\pm 0.003$ & $0.699\pm 0.003$ \\
 & 10 & $14.910\pm0.008$ & $0.683\pm 0.003$ & $0.699\pm 0.003$ \\
 & 20 & $7.45\pm0.02$ & $0.694\pm 0.003$ & $0.688\pm 0.003$ \\
\hline
\end{tabular}
\caption{Results for parameters $\tau$, $\beta_1$, and $\beta_2$ from the numerical fittings. Each value is an average over allowed values of $\epsilon$ for each $f_r$.}
\label{table:one}
\end{table}

We conclude that the technique of ABWV is more robust to slow drifts than WVA, and it also offers background subtraction. As a result, the technique gives one the possibility for longer collection times and larger signal-to-noise ratios. 

We proceed now to improve the state-of-the-art WVA result for velocity measurements. By averaging 78 pulses in a similar configuration to Fig.~\ref{fig:setup}, the best reported averaged velocity in Ref.~\cite{Velocimetry} was of 400$\pm$400 fm/s. 
Such a result gives a sensitivity of $\sim3.5$ pm/s per averaged pulse when using WVA. In our case, using ABWV, we set a frequency of $f_r=$2.5 mHz, a voltage on the piezo of $V_{pp}=$0.3 mV, and a time constant of $\tau=67$ s. We used an average value of $\epsilon=110$ mrad for 358 collected pulses and obtained a estimated average velocity of 380 $\pm$ 60 fm/s. The obtained sensitivity was of $\sim1.1$ pm/s per averaged pulse. Note that 358 pulses at a repetition rate of 2.5 mHz corresponds to a total acquisition time of about 40 hours\footnote{These pulses were collected in 6-hour daily sets taken during one week. Each of the sets was taken during night time to avoid external high-frequency vibrations.}
. As a merit of comparison, our precision in velocity corresponds to a Doppler shift in the laser light of $2\Delta v/\lambda\sim150$ nHz. Measuring such a shift in a standard continuous-wave Homodyne configuration would produce a beat note of about 11 weeks!


\section{Conclusions}\label{section conclusions}
We have compared the recently-proposed technique of using two detectors to subtract the signals produced by two almost-equal weak values (ABWV) to the better known technique of amplification due to one large anomalous weak value (WVA). We performed precision measurements of linear velocities of one of the mirrors in a Michelson interferometer, and further expand the already-known advantages of the ABWV technique over WVA. We confirm that practical advantages of balancing signals make the ABWV technique more robust against slow drifts and systematic errors than the WVA protocol. 

The technique of ABWV offers larger signal-to-noise ratios than WVA, however the well-behaved range for the parameter $\epsilon$ extends only up to $\sim0.2$ rad. On the other hand, the technique of WVA breaks down for angles below $\sim0.2$ rad, but it offers good performance for angles above it. Small values of $\epsilon$ are always desired in weak-values metrological techniques. However, we have found here that ABWV and WVA complement each other to allow precision measurements for almost any given value of $\epsilon$ in the range from 10 to 800 mrad (see Fig.~\ref{fig:shifts}).

We were able to measure a Doppler shift of $\sim$950 nHz with a precision of $150$ nHz after 40 hours of integration time. We note that the competitive technique of continuous-wave Homodyne detection would require at least 11 weeks to resolve the corresponding beating signal.

Experimental demonstrations of the ABWV technique have been done for measurements of time shifts in pulses~\cite{ABWV,Weitao}, as we do here. The technique of WVA has successfully been used in configurations that require split detection to track the field's transverse distribution of a continuous-wave laser beam. As a future work, it would be interesting to evaluate the performance of the ABWV technique for such cases, where two high-resolution beam-profile detectors would be required. 



\section*{Acknowledgements}
This work was funded by the Army Research Office (Grant No. W911NF-12-1-0263), Northrop Grumman Corporation, and the Department of Physics and Astronomy at the University of Rochester. 


%

\end{document}